# Photo-ionization structures of Planetary Nebula IC 2003 with [WR] central star

K. Khushbu [a,b,*], C. Muthumariappan [a]

[a] *Indian Institute of Astrophysics, Koramangala, Bangalore 560 034, India*
[b] *Pondicherry University, R.V. Nagar, Kalapet 605014, Puducherry, India*

**Abstract**

We present ionization structures of IC 2003, a planetary nebula with [WR] central star, using a 1-D dusty photo-ionization code: "*CLOUDY 17.03*". The photo-ionization model is constrained by archival UV emission line fluxes, medium-resolution optical spectroscopy, *IRAS* 25 $\mu m$ flux, absolute $H\beta$ flux, and the mean angular size of the nearly spherical optical nebula. To constrain the carbon abundance and the effect of photo-electric heating in the ionized gas, we used UV emission lines. We considered an amorphous carbon dust grain with MRN and KMH size distributions to address the importance of photo-electric heating in the ionized nebula. We show that KMH grain size distribution with quantum dust heating reproduces the observations quite well. We construct the ionization structures of different elements at their different ionization stages in the nebula. We derive the physical properties of the planetary nebula and its chemical composition, as well as the parameters of its central star. The estimated nebular dust-to-gas mass ratio is $2.37 \times 10^{-3}$, and the enhanced photo-electric heating yielded by small dust grains is 9.4% of the total heating. We considered the H-poor model atmosphere for the central star; the effective temperature of the central star is 177 kK, the specific gravity log(g) is 6, and its luminosity ($L_*$) is 6425 $L_\odot$. The derived central star parameters plotted on stellar evolutionary tracks correspond to a central star mass of 0.636 $M_\odot$ and to a progenitor mass of 3.26 $M_\odot$.

## 1. Introduction

Planetary Nebulae (PNe; in singular: PN) form during a brief phase of the evolution of low- and intermediate-mass stars (∼0.8-8$M_\odot$) after their asymptotic giant branch (AGB) phase of evolution before they attain white dwarf configurations. The Hong-Kong/AAO/Strasbourg/Hα PNe (HASH) database by Parker et al. (2016) has listed a total of 3795 PNe present in our galaxy. PN plays an important role in studying stellar evolution, galactic chemical evolution, and nucleosynthesis processes in AGB stars (Ventura et al., 2017; Kwitter and Henry, 2022). The central stars of PNe are quite hot (effective temperature of the central star ($T_{eff}$) above 30 kK), and the hard-UV radiations of the central star can ionize the surrounding gas. The ionized envelopes emit several bright spectral lines (like [OIII] at 5007 Å and [NII] at 6584 Å), which are important for the cooling of PNe. The physics and the chemical compositions of the nebula can be probed in detail using these emission lines, and they are important to understand the thermal and ionization structures of PNe and the evolutionary history of their progenitors. Spectral lines arising from different ionization stages of elements are used to derive nebular abundances, electron temperatures and densities by using empirical methods (Kingsburgh et al., 1994). Though the simple empirical methods are widely being employed to derive the abundances and nebular diagnostics, they are not very accurate; a reason for this is that the determination of elemental abundances other than He requires accurate determination of electron temperature, which has a distribution across the nebula. Secondly, the ionizing photons from the central star are intercepted by the grains in the nebula, and gas and dust strongly compete with each other in absorbing the stellar UV photons (Lyman, 1948; Dopita and Sutherland, 2000; van Hoof et al., 2004). The presence of grains in the nebula shrinks the size of the ionized gas and causes a notable decrease in the nebular H$\beta$ flux while increasing its IR continuum (Stasińska and Szczerba, 2001). Also, grain physics follows a non-linear functionality with charge and size distribution when the grain sizes are smaller (smaller than $\sim$ 50 Å). Emission lines from highly ionized species such as [OIV] 1402 Å, [CIV] 1549 Å, and [CIII] 1909 Å are primarily affected by the presence of grain heating (van Hoof et al., 2004). It is therefore important to construct detailed thermal and ionization structures of PNe using tailored dusty photo-ionization models, giving gas and dust to play their due role in the radiative transport. This helps to constrain accurately the chemical and physical parameters of PNe as well as to derive the central star parameters (Maciel et al., 2010; García-Rojas et al., 2016). Photo-ionization models were derived for some PNe involving gas and dust components earlier by different authors (van Hoof and van de Steene, 1999; Bohigas, 2008; Otsuka et al., 2017; Bandyopadhyay et al., 2020) The central stars of PNe (CSPNe) are broadly classified into two groups based on their hydrogen content, viz. the hydrogen-rich and the hydrogen-poor CSPNe (Acker and Neiner, 2003; Weidmann and Gamen, 2011), and also by Muthumariappan and Parthasarathy (2020) (hereafter, MP20). About 15% of the CSPNe are known to show the Wolf-Rayet phenomena (hereafter [WR] stars, (Acker and Neiner, 2003)). They are similar to the massive WR stars, showing fast stellar winds with strong and broad emission lines of helium and highly ionized species such as carbon and oxygen (Blöcker, 2001; Acker and Neiner, 2003; Werner and Herwig, 2006). The origin and the evolutionary status of [WR] stars are largely unknown, and PNe around these stars ([WR]PNe) have the evolutionary history imprinted with them, as discussed by MP20. One such property is the large radial temperature gradient (Stasińska and Szczerba, 2001) for [WR]PNe and the effective photo-electric heating of the nebula by the very small grain population, which can be significantly larger for [WR]PNe in comparison with the normal PNe, as suggested by MP20. A detailed, dusty photo-ionization modelling of [WR]PNe will confirm this and, in turn, will help us to understand the differences in the physical and chemical structures of [WR]PNe in comparison to the normal PNe, if any.

* Corresponding author.
*E-mail addresses:* khushbu@iiap.res.in, khushbukathpalia97@gmail.com (K. Khushbu).

Such a detailed study has not been done for many objects, and here we intend to make a complete 1-D dusty photo-ionization model of a [WR]PN IC 2003.

IC 2003 is a small and faint PN located in the Perseus constellation, which consists of a bright disk with a small blob of high surface brightness (Aller and Czyzak, 2003). The central star of the PN is [WR] nature with WC3 subtype and OVI sequence with $T_{eff}$ of 112 kK (Feibelman, 1997; Aller and Walker, 1970). This PN belongs to an excitation class of 9 as found by Gurzadyan and Egikyan (1991), also by Muthumariappan et al. (2022) (hereafter MAS22). The nebula has a carbon-rich chemistry, as discussed in MAS22. IC 2003 is located at a distance of 4.33 kpc (Frew et al., 2015) and has a mean angular diameter of 15.44″. The photo-ionization model of IC 2003 was worked out earlier by Koeppen (1983), and MAS22 presented a 1-D dusty photo-ionization model with a black-body approximation for the central star spectrum. Their models could not reproduce the observations very well, and the photo-electric heating needed to be better addressed. MAS22 suggested to include a more realistic nature of the stellar input spectrum. Also, it proposed to work out grain heating with different size distributions in the nebula for better reproduction of the observations, which we intend to do here. The abundances derived from the models can also have a significant impact from these considerations. In this paper, we follow the basic methodology of MAS22 for photo-ionization modelling but revisit the models by including an accurate model atmosphere for the central star (in contrast with the black-body approximation considered earlier) and address the scale of grain heating with two different dust size distribution functions. Subsequently, we aim to study the importance of grain heating in the determination of nebular abundances and how the values of stellar parameters vary with these new inclusions in the models. We search for a best-fit photo-ionization model for IC 2003 to derive the thermal and ionization structures of the PN and to determine the accurate parameters of CSPNe. We use emission line fluxes of the UV/Optical domain and IR broad-band photometric observations to constrain the proposed model effectively.

We organize the paper as follows: Section 2 describes the details of the observations and data analysis methods. Section 3 describes the empirical analysis of abundances using *NEAT* and discusses the methodology of our 1D Dusty photo-ionization modelling incorporating the stellar model atmosphere and different grain size distributions. In Section 4, we present the results obtained from different models we have worked out with and without considering the quantum heating component, discuss their implications, and propose the most suitable model for IC 2003. We also present here the nebular and central star parameters and the ionization structures of the PN derived from the best-fit model. Section 5 refers to the concluding remarks of the study.

## 2. Observations and data reduction

We conducted medium-resolution grating spectroscopic observations on IC 2003 in the optical domain using a low-intermediate resolution OptoMechanics Research (OMR) spectrograph attached with the f/13 Cassegrain focus of the 2.34 m Vainu Bappu Telescope located at VBO Kavalur, Tamilnadu (India) (Prabhu et al., 1998). The spectrograph has a 25 mm long slit (which subtends $\approx 2.8'$ in the sky) with a slit-width of 300 microns ($\approx 2''$ in the sky). PNe are usually spatially extended, axisymmetric objects; hence to have a better representation of the full axisymmetric nebula, we added the spectra taken at two distinct slit position angles of 0° and 270°. The slit was set passing through the centre of the nebula and covering the whole ionized gas, which is a required condition for our modelling. We used a 600 lines/mm grating with a spectral resolution of 1000 (2.9 Å/pixel at 5000 Å) for two distinct wavelength regions, viz. blue region and red region, to cover the spectral range of 3700 to 8400 Å. The blue region covered a spectral range of 3700 Å to 6300 Å, and the red region has a spectral range of 5800 Å to 8400 Å. Multiple frames with different exposure times for each setting were taken at these slit positions, as given in Table 1. The data was recorded using a Peltier-cooled Andor CCD detector with 1024 × 256 pixels rectangular array format with a pixel size 26 $\mu$m (Andor, 2022). The detector was cooled to the temperature of 178K using a coolant. We have taken the bias, dark, and flat frames during the target observations with the same instrument settings to calibrate CCD. FeAr and FeNe comparison lamp frames were taken for wavelength calibration for the blue and red regions, respectively. Also, spectro-photometric standard HD 19445 (Oke and Gunn, 1983) was observed for the flux calibration with an exposure time of 10 minutes for each wavelength setting.

Further, we have used UV emission line fluxes and IR continuum fluxes of IC 2003 from archive for this study, along with our optical spectra. We have taken eight UV line fluxes of *IUE* (1549 Å - 2424 Å) given in Wesson et al. (2005) which includes major cooling lines of carbon ([CIV] 1549 Å and [CIII] 1909 Å). *IRAS* 25 $\mu$m band flux was taken from the NASA IPAC IR science archive to trace continuum dust. We excluded the flux at 12 $\mu$m band of *IRAS* as PAH emission features contaminate this band and the flux at 60 $\mu$m band was not considered as this data quality was poor. We used the standard procedure of spectroscopic data reduction using the *IRAF* packages (Valdes, 1994). The CCD correction includes bias subtraction and flat-field correction. Spectrum was extracted from the CCD corrected frames and sky subtracted and then subjected to the wavelength and flux calibrations. The wavelength calibration was done using lamp frames by identifying several lines and fitting a polynomial. We used the 5875 Å HeI line, which is present in both the blue and red parts of the spectra, to scale one with respect to the other. We then combined the spectra taken at two slit positions into one spectrum, which averaged the emission line flux variation across the nebula as well as enhanced the net signal-to-noise ratio. The spectrum was then corrected for the interstellar reddening using the logarithmic interstellar reddening parameter using Balmer lines $c_{H\beta}$. To calculate $c_{H\beta}$, we used the relation from Osterbrock and Ferland (2006) given below:

$$\frac{I_{H\alpha}}{I_{H\beta}} = \frac{I^0_{H\alpha}}{I^0_{H\beta}} 10^{-c_{H\beta}[f(H\alpha)-f(H\beta)]} \qquad (1)$$

where $I^0_{H\alpha}/I^0_{H\beta} = 2.86$ (theoretical value for a hydrogen-rich ionized nebula), $I_{H\alpha}/I_{H\beta}$ is the observed ratio, and f($\lambda$) is the extinction law for each wavelength at standard reddening curve R = 3.1 within the galaxy (Osterbrock and Ferland, 2006). We calculated $c_{H\beta}$= 0.4, which is consistent with the values given in the literature (for example, $c_{H\beta}$ = 0.42, (Kingsburgh et al., 1994). We used this value to correct for reddening by applying the interstellar extinction correction relation from Fitzpatrick (1999). Flux calibration was done using wavelength-calibrated spectroscopic standard. Then, we measured emission line fluxes at each line in the de-reddened flux calibrated spectra by fitting a Gaussian to it. The measured line fluxes are listed (relative to the flux at $H_\beta = 100$) in Table 3.

## 3. Methodology

The de-reddened emission line fluxes from our optical spectra were used to derive the ionization structures and the parameters of the PN IC 2003 and its central star. UV lines are used to constrain the photo-electric heating accurately, and the *IRAS* continuum flux at 25 $\mu$m is used to constrain the cooling by classical grains. We compute the diagnostics of the nebula using the empirical method and then invoke photo-ionization modelling.

### 3.1. Empirical Nebular diagnostics using NEAT

We have used the publicly available Nebular Empirical Analysis Tool (NEAT, Wesson et al. (2012)) to estimate the nebular diagnostics of IC 2003 using optical and UV emission line fluxes. NEAT identifies the ionic species using the rest wavelengths of emission lines listed in its atomic data directory and derives the nebular electron temperature ($T_e$) and electron density ($n_e$) and the empirical abundances of elements at their different ionization stages. Abundances of the unseen ions were estimated by NEAT using the in-built ionization correction factors (ICFs). Table 4 shows the nebular abundances of IC 2003 derived using this method. The empirically derived nebular abundances are reasonably approximate, and they are very useful initial values for any photo-ionization models. It reduces the computation time taken to arrive at the best-fit model quite significantly. Below, we describe the method of photo-ionization modelling of IC 2003, which we followed using the nebular empirical abundances.

Table 1
Optical spectroscopic observation Log.

| Slit P.A(°) | Spectral Range | Exposure(minutes) | Observing Date |
|---|---|---|---|
| 0 | Blue | 45 | 11/01/22 |
| 0 | Red | 15 | 11/01/22 |
| 270 | Blue | 35 | 12/01/22 |
| 270 | Red | 35 | 12/01/22 |

### 3.2. 1D Dusty photo-ionization model

We used 1-D dusty photo-ionization code *CLOUDY* (*version* : 17.03; Ferland et al. (2017)) to model the photo-ionization structures of the [WR] PN IC 2003. The basic methodology followed for our modelling was discussed in MAS22. The distance to the nebula is a critical parameter as the measured nebular emission line fluxes, and stellar luminosity are inversely proportional to the square of the distance. Hence, we ensured that this parameter was well measured in the literature, and we treated this as a fixed parameter. The distance to PN IC 2003 was taken from Frew et al. (2015), which has an accuracy better than 30%. Our present study includes three significant improvements after the basic model published by MAS22. They are: 1) the central star spectrum of the PN is more accurately described by the H-poor stellar model atmosphere, as the black-body spectral energy distribution (SED) deviates significantly in the UV to the extreme-UV region where the SED of CSPN peaks. Hence, black-body approximation only gives a rough estimate of the nebular diagnostics and abundances (Armsdorfer et al., 2001). 2) photo-electric heating provides a significant contribution to the thermal balance of the ionized nebula, and this effect can be better measured using the UV fluxes sensitive to grain heating. 3) as discussed in MAS22, the MRN size distribution of the grain model does not provide a significant amount of the very small grain population, which is too important to heat the nebula through the photo-electric effect. Hence, the most appropriate grain size distribution for the PN must be worked out. Moreover, carbon is a major cooling agent of the nebula, like oxygen and nitrogen, which does not have any strong emission line in the optical domain. UV has strong carbon lines, and fluxes at these lines help to determine the abundances and thermal structures of the nebula better. We discuss below the new inclusions made in the models in detail.

#### 3.2.1. Stellar model atmosphere

The SED of CSPNe in the UV to extreme-UV region is strongly modulated due to the metallic opacity of the atmosphere, which absorbs a significant amount of radiation in this wavelength domain. For a given stellar temperature, the black-body model gives more energy in UV to the extreme-UV region than the model atmosphere, and they deviate significantly in this region (see Fig. 1). Hence, to describe the central star of IC 2003, we used the 'table star Rauch PG1159' H-poor stellar SED grid of models computed by Rauch (2003). Rauch PG1159 grids of the model create non-local thermal equilibrium (non-LTE), plane parallel line blanketed stellar atmospheres containing a set of models with $T_{eff}$ from 40 kK to 190 kK and log g = 5 to 9. We manually varied log g and $L_*$ within the given range. For a fixed $T_{eff}$, we find that $L_*$ is not sensitive to log g. Therefore, we fixed log g at 6.0 based on the best fit obtained by $T_{eff}$. The model atmosphere we have chosen is well suited for the H-deficient central stars of PNe with He atmosphere, like [WR] CSPNe.

#### 3.2.2. Grain nature and its size distribution

For a given grain nature and the grain size distribution, the refractive index file available with *CLOUDY* and the ISM dust-to-gas mass ratio ($2.37 \times 10^{-3}$ estimated from Table 4 of van Hoof et al. (2004)) is used to create the nebular opacity file, which *CLOUDY* uses to compute the radiative transfer. The opacity file has the values of the opacity of the gas and dust components of the nebula at different wavelengths for the ISM dust-to-gas mass ratio. The presence of PAH features in the *Spitzer-IRS* spectra of IC 2003 clearly shows a carbon-rich circumstellar environment, which is also supported by the carbon-to-oxygen abundance ratio (C/O) measurements by several authors (Koeppen, 1983; Kingsburgh and Barlow, 1994). We have considered the grain type to be amorphous carbon as the *Spitzer-IRS* spectrum of this object does not show the features of crystalline dust. Further, in the dense AGB atmosphere, grain condensation favours the amorphous form of dust. To test the most suitable size distribution for the grains in the nebula, we explored models with MRN and KMH grain size distributions and compared the model fluxes with observations. The MRN grain size distribution is described as a power-law as follows (Mathis et al., 1977):

$$\frac{dn(a)}{da} \propto a^{-3.5} \tag{2}$$

where n(a) is the number density of grains with size a, da is the bin size, and the power-law index is 3.5. MRN distribution can be obtained using *CLOUDY* by simply choosing the minimum and maximum grain sizes, $a_{min}$ and $a_{max}$ respectively, and the power-law index. The KMH distribution has the same power-law pattern for small grains with an exponential decay for larger grains (Kim et al., 1994). In *CLOUDY*, the general form of power-law size distribution with exponential cutoff is given as follows (Ferland et al., 2017):

$$\frac{dn(a)}{da} \propto a^{-3.5} F(a;\beta) C_l(a;a_l,\sigma_l) C_u(a;a_u,\sigma_u) \tag{3}$$

where $dn/da$ is the number density slope of the extinction curve, a is the grain size, $F$ gives the extra curvature, and $C_l$ and $C_u$ are the functional limits for exponential cutoff with $a_l$ and $a_u$ as the lower and upper limits and $\sigma_l$ and $\sigma_u$ are the scaling factors respectively. To get the desired

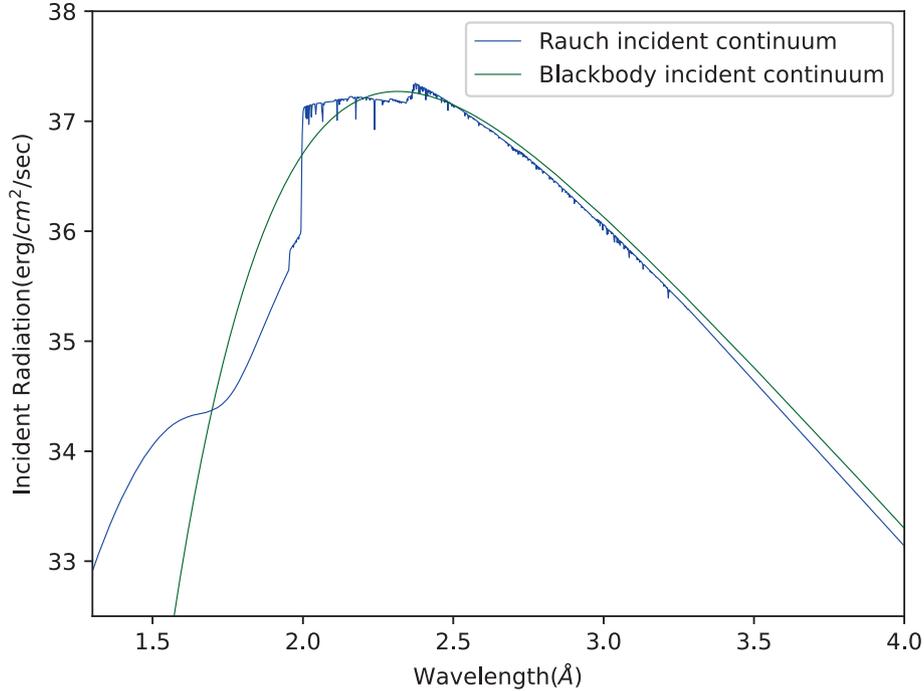

Fig. 1. Incident Continuum of the central star of IC 2003 for black-body and Rauch model atmosphere.

KMH distribution, we set $\beta = 0$ leading $F(a;\beta) = 1$. The lower exponential cutoff is not required; hence, $C_l = 1$ (by setting $a_l = 0\mu m$). The upper exponential cutoff is described as

$$C_u(a; a_u, \sigma_u) = \begin{cases} 1 & if\ a \leqslant a_u \\ exp[-(a-a_u)/\sigma_u] & if\ a > a_u \end{cases} \quad (4)$$

The required KMH size distribution can be obtained using this, and it has the following form:

$$\frac{dn(a)}{da} \propto a^{-3.5} e^{(a-a_u)/\sigma_u} \quad (5)$$

with the appropriate values set for $\sigma_u, a_{min}, a_u$, as required for our study. Though the default power-law index is 3.5 for MRN and KMH grain size distributions; we have also worked on the best-fit models considering a power-law index of 3.7.

We prepared the opacity files for MRN and KMH size distributions by resolving the grain sizes with 31 bins in the logarithmic scale. For a given dust mass, the KMH distribution puts more mass in the very small grain population with grain size $a \leqslant 50Å$, as compared to the MRN size distribution, and hence a larger projected area for the grains. This is required to determine the quantum heating effectively and for computing the photo-ionization structure of PN (van Hoof et al., 2004). With this, we can model the thermal balance of a PN with or without considering quantum heating to examine the scale of its importance.

Empirically derived values of electron density $n_e$ and elemental abundances from NEAT were taken as the initial values of input parameters. The electron density is assumed to be constant within the ionized gas as done earlier by Bohigas (2008). The empirical values are reasonably approximate, and hence, they are considered as semi-free parameters in modelling and varied in a given narrow range. Other parameters such as the stellar temperature ($T_*$) and luminosity ($L_*$), the nebular parameters like the inner radius ($R_{in}$), outer radius ($R_{out}$), filling factor (ff), and the dust parameters such as mean dust temperature ($T_d$) and dust-to-gas mass ratio ($m_d/m_g$) are considered as free parameters and varied within their possible ranges.

We optimized the model fluxes of seven emission lines relative to H$\beta$, viz. [OIV] 1402 Å, [CIII] 1909 Å, [OIII] (4363 & 5007 Å), HeII 4686 Å, HeI 5875 Å and [NII] 6584 Å against their observed values. The maximum error in the observed fluxes is 5%, and hence, $CLOUDY$ was set to take this as the maximum allowed deviation while modelling these line ratios. UV lines are optimized to constrain the scale of quantum dust heating, and HeII lines are used to constrain $T_{eff}$ and He abundance. The thermal structure of the nebula was optimized using the major cooling lines of [OIII], [CIII] and [NII]. We optimized the absolute $log(H\beta)$ intensity and the angular extend of the H$\beta$ emitting region (Stromgren sphere) with their respective observed values. We also optimized the $IRAS$ continuum photometric flux at its 25 $\mu$m band. The maximum deviation that $CLOUDY$ can have while reproducing the H$\beta$ flux, angular diameter and the 25 $\mu m$ $IRAS$ band flux are set respectively at 5%, 10% and 20%. The maximum possible error derived that is sensed by $\chi^2$ for any parameter is 10%. The optimization provides a net $\chi^2$ value for a given model, and the model that corresponds to the minimum $\chi^2$ value was taken as the best-fit model.

## 4. Results and discussion

We present here the results obtained from our 1-D dusty photo-ionization models of the PN IC 2003, considering grains as an important component in determining the thermal and ionization structures of the nebula. We analyse the models derived for different possible grain size distributions, viz. MRN & KMH size distributions. In addition, we examine the importance of the grain heating effect by working out models with and without considering grain heating. These models are presented, and the best-fit model is discussed in detail below.

### 4.1. KMH non-heating model (model 1)

The general form of the KMH grain size distribution function is given in Eq. 5 of Section 3.2.2. The function has a power-law distribution up to a grain size of $a_u$, and then it follows an exponential decay. We varied $C_u$ by adjusting the parameters $a_u$ and $\sigma_u$ until we found the most suitable value; we find that $a_u$ and $\sigma_u$ should have their respective values of 0.02 $\mu m$ and 0.2 $\mu m$ in order to get an acceptable dust-to-gas mass ratio. We opted for lower and upper grain size limits of $a_{min}=$ 0.0025 $\mu m$ (as we need to keep the size as small as possible to get more photoelectric energy, however grains with smaller than 25 Å are not expected to survive in the hostile environment) and $a_{max}=$ 0.5 $\mu m$ (since $a_{max}$ does not cause a significant change when varied for different possible values, it is fixed at the value typically used for PNe), to account for the very small (size ⩽100 Å) as well as classical (size ⩾100 Å) grain populations in the PN. The classical grains of size 100 Å to 0.5 $\mu m$ can attain thermal equilibrium as the heating by UV radiation is balanced by cooling by IR emission at any instant; they are at a temperature of ≈ 100 K. Very small grains with sizes 25 Å – 100 Å are transiently heated to very high temperatures (up to 1000 K) by absorbing a single UV photon and cool down relatively slowly by IR emission before the next absorption (van Hoof et al., 2004; Muthumariappan, 2017). Hence, they are not at the equilibrium temperature. These small grains are also responsible for emitting photoelectrons and heating the gas. *CLOUDY* treats and computes their statistical temperature at every radial distance.

The opacity file for gas and dust was created for KMH grain size distribution for the ISM dust-to-gas mass ratio. The *CLOUDY* code subsequently called this file to determine the nebular photo-ionization structures without considering the dust quantum heating mechanism. The dust-to-gas mass ratio of the PN was scaled with respect to the ISM using a scale factor to fit the models. The free parameters were varied in steps within their possible values, and the semi-free parameters were varied within the range obtained from the empirical methods. Several possible models were computed for different sets of input variables. These models were examined and optimized against the observations using the inbuilt *CLOUDY* optimization method, giving a value of net $\chi^2$ for each model, which measures the overall fit of the data. The best-fit model was chosen from the minimum $\chi^2$ value. Table 3 lists the line fluxes reproduced for this model and compared with the observations, and the parameters of this model are listed in Table 2 ($KMH_{nh}$ column), including the $\chi^2$ value for each category of observed parameters. As is seen from Table 3, the carbon emission line fluxes in UV are not reproduced satisfactorily with the observations, which is possibly due to the fact that the quantum heating is switched off for this model. The elemental abundances derived for this model are given in Table 4 and are also compared with the *NEAT* abundances. The empirical value of C/O is a little less than unity, showing an oxygen-rich (O-rich) environment. However, the presence of PAH features in the *Spitzer*-IRS spectrum clearly shows the envelope to be carbon-rich (C-rich). We can see from Table 4 that the model values of nitrogen-to-oxygen abundance ratio (N/O) and C/O are higher compared to their empirical values. The model value of C/O shows the expected C-rich envelope for IC 2003, and the empirical values are hence not very accurate. The He abundance relative to H derived from this model (0.2, see Table 4) is quite high. The value of $T_{eff}$ indicates that the central star of PN IC 2003 is very hot, but the high He abundance suggests that this parameter may be overestimated. This also could be related to the absence of quantum heating in this model. Fig. 1. shows the radial variation of electron temperature for IC 2003 from its inner edge as predicted by the KMH no-heating model. The radial temperature gradient is not high for this model, as inferred from the model UV fluxes fitted to their observed values.

### 4.2. MRN Heating Model (model 2)

We considered the MRN grain size distribution for this model, which has a general power-law form as given in Eq. 2 (Section 3.2.2). The opacity file for gas and dust was computed for the ISM dust-to-gas mass ratio. The minimum grain size for this distribution $a_{min} = 0.0025$ $\mu m$, and the maximum grain size $a_{max} = 0.5$ $\mu m$. Several models were computed by *CLOUDY* for different possible sets of input variables, incorporating the grain quantum heating in addition to the photo-ionization heating of the nebula. The models were assessed using $\chi^2$ method, and the model with minimum $\chi^2$ is presented here. Table 3. lists the model emission line fluxes and shows that the overall reproduction of the observed emission line fluxes are improved from model 1. The best-fit model parameters are shown in Table 2, and Table 4 lists the derived abundances. As seen from these tables, the central star temperature and the He abundance are relatively lower in this model when compared to their values in the KMH non-heating model. However, the effective temperature of the central star is higher than the value derived using the black body approx-imation of the MRN model presented by MAS22 (refer to their Table 4). This shows the significance of considering a realistic model atmosphere for the hot [WR] central star over a Planckian approximation to define the input SED of the PN. Table 2 gives the grain-to-gas heating ratio (GrGH), which is only 5.5%. This is the ratio of the thermal energy acquired from photo-electric heating to the energy from photo-ionization heating by the ionized gas. This is defined by the net projected surface area of grains for a given volume (indicated by $<a^2/a^3>$ in Table 2) of the very small grain population defined by the MRN size distribution function. As can be seen in Table 4, the N/O and C/O ratios are significantly larger in model 2 compared to their empirical values (and also larger than in model 1), asserting the C-rich nature of the nebula. The He abundance is also reduced significantly compared to its value in model 1. However, the absolute $H\beta$ flux predicted by model 2 is lower than the observed value. The overall model fit to the data, as judged by the net $\chi^2$ value, is better than for model 1. though it still may not be the best possible one as the UV lines do not fit well. The outward radial electron temperature profile for IC 2003 corresponding to the MRN heating model is shown in Fig. 1. The radial temperature gradient is not significantly different from that of model 1, and it is not high enough, as inferred from the model UV fluxes fitted to their observed values.

Table 2
Parameters of IC 2003 derived from different models viz., KMH non-quantum heating model(KMH_nh), MRN model with quantum heating(MRN_h), KMH model with quantum heating(KMH_h), MRN model with quantum heating with power law index 3.7(MRN_h1) and KMH model with quantum heating with power law index 3.7(KMH_h1). See the text for more details.

| Model Parameters | KMH_nh | MRN_h | KMH_h | MRN_h1 | KMH_h1 |
|---|---|---|---|---|---|
| $T_{eff}$ (K) | 189,000 | 186,000 | 178,000 | 174,121 | 178,619 |
| $L_*(L_{sun})$ | 6,300 | 6,100 | 6,400 | 6,150 | 6,023 |
| $log(g)$ | 6.0 | 6.0 | 6.0 | 6.0 | 6.0 |
| $D(kpc)$ | 4.33 | 4.33 | 4.33 | 4.33 | 4.33 |
| $R_{in}(cm)$ | $1.65 \times 10^{17}$ | $1.11 \times 10^{17}$ | $1.23 \times 10^{17}$ | $1.21 \times 10^{17}$ | $1.39 \times 10^{17}$ |
| $R_{out}(cm)$ | $5 \times 10^{17}$ | $5 \times 10^{17}$ | $5 \times 10^{17}$ | $5 \times 10^{17}$ | $5 \times 10^{17}$ |
| $\epsilon$ | 0.606 | 0.602 | 0.680 | 0.601 | 0.575 |
| $n_e(cm^{-3})$ | 1,737 | 1,721 | 1,628 | 1,730 | 1,769 |
| $Diameter(″)$ | 15.44 | 15.44 | 15.44 | 15.44 | 15.44 |
| (Observed) | 15.78 | 15.78 | 15.78 | 15.78 | 15.78 |
| $F_{H\beta}$ | $4.22 \times 10^{-11}$ | $4.15 \times 10^{-11}$ | $4.17 \times 10^{-11}$ | $4.15 \times 10^{-11}$ | $4.20 \times 10^{-11}$ |
| (Observed) | $4.17 \times 10^{-11}$ | $4.17 \times 10^{-11}$ | $4.17 \times 10^{-11}$ | $4.17 \times 10^{-11}$ | $4.17 \times 10^{-11}$ |
| $<a^2/a^3> (cm^{-1})$ | $4.6 \times 10^5$ | $2.8 \times 10^5$ | $4.6 \times 10^5$ | $4.3 \times 10^5$ | $6.4 \times 10^5$ |
| $m_d/m_g$ | $2.37 \times 10^{-3}$ | $2.37 \times 10^{-3}$ | $2.37 \times 10^{-3}$ | $2.54 \times 10^{-3}$ | $2.52 \times 10^{-3}$ |
| $T_{dust}(K)$ | 74 | 74 | 74 | 74 | 74 |
| GrGH | – | 5.5 | 9.4 | 8.9 | 11.1 |
| $\chi^2$ (Relative flux) | 14.31 | 10.45 | 8.105 | 4.536 | 4.218 |
| $\chi^2$ (Absolute flux) | 0.055 | 0.0086 | 0.0001 | 0.0005 | 0.029 |
| $\chi^2$ (Angular diameter) | 0.049 | 0.049 | 0.048 | 0.049 | 0.049 |
| $\chi^2$ (IR Photometry) | 0.168 | 0.0118 | 0.002 | 0.0084 | 0.064 |
| Net $\chi^2$ | 14.58 | 10.63 | 8.15 | 4.60 | 4.36 |

### 4.3. KMH heating model (model 3)

With the same grain size distribution and the opacity file of gas and dust adopted for model 1, we included the process of quantum heating in the nebula by the very small grain population for the KMH heating model (model 3). We computed several possible models by varying the input variables within their given ranges. These models were optimized as done before, and the best-fit model with the lowest $\chi^2$ is reported here. As can be seen in Table 3, the fluxes of the high ionization UV lines of [OIV] (1401 Å), [CIV] (1549 Å) and [CIII] (1909 Å), which trace the grain heating of the nebula (van Hoof et al., 2004), are better reproduced in this model in compared to model 2. The emission line flux for [CIV] 1549 Å is 10% enhanced and 7.98% for [CIII] 1909 Å in model 3 over the non-heating models. The net $\chi^2$ value of model 3 is lower than the values of model 1 and model 2., implying the best reproduction of the overall observations by model 3.

The grain heating to the gas heating ratio (GrGH) is 9.4% for this model, which is larger than for model 2 and is quite significant. The quantum heating effect is highest for the KMH grain size distribution, which is due to the fact that the KMH grain size distribution populates the very small grain population more than the MRN size distribution. This is indicated by the factor $<a^2/a^3>$ which is $\sim$ 60% higher for the KMH distribution than the MRN distribution, showing a larger projected mean surface area for the stellar UV radiation field as described in Section 3.2.2, which in turn will substantially increase the grain-gas heating (GrGH) effect. With a power-law exponent of 3.7, both MRN and KMH give better GrGH values than their respective models with a power-law index of 3.5 (see Table 2 for the model parameters). The KMH model makes a better fit to the data than the MRN model even if we take a different power-law index. However, we choose the models produced with the default power-law index of MRN and KMH distributions (3.5), which is expected and typically being used in PNe studies (van Hoof et al., 2004), for comparison. From this we propose model 3 as the most suitable model for IC 2003 and consider its parameters for our further analysis.

Table 3
Comparison of observed values with modelled emission line fluxes and *IRAS* continuum flux for different models.

| Wavelength (Å) | Line ID | Observed | KMH_nh | MRN_h | KMH_h |
| --- | --- | --- | --- | --- | --- |
| 1401.00 | [O IV] | 20.80 | 16.13 | 16.76 | 17.31 |
| 1549.00 | [C IV] | 353.16 | 330.91 | 452.07 | 367.80 |
| 1640.42 | He II | 372.09 | 412.65 | 437.69 | 372.40 |
| 1663.00 | [O III] | 23.37 | 29.57 | 32.74 | 28.28 |
| 1751.00 | [N III] | 22.34 | 15.05 | 16.12 | 17.61 |
| 1909.00 | [C III]b | 420.54 | 358.79 | 381.32 | 389.94 |
| 2325.00 | [C II] | 21.97 | 19.25 | 21.89 | 18.37 |
| 2424.00 | [Ne IV]b | 34.49 | 117.4 | 122.71 | 119.58 |
| 3868.75 | [Ne III] | 131.50 | 203.92 | 209.12 | 211.05 |
| 3887.44 | He II | 23.38 | 0.63 | 0.61 | 0.59 |
| 3893.52 | O II | 23.59 | | | |
| 3967.46 | [Ne III] | 63.80 | 62.18 | 63.68 | 64.38 |
| 4101.74 | H I | 47.78 | 24.07 | 24.31 | 24.06 |
| 4340.47 | H I | 54.22 | 46.53 | 46.78 | 46.45 |
| 4363.21 | [O III] | 16.73 | 21.29 | 21.23 | 20.80 |
| 4469.00 | He I | 2.23 | 3.28 | 3.39 | 3.47 |
| 4539.00 | He II | 1.86 | 2.79 | 2.66 | 2.61 |
| 4685.68 | He II | 58.68 | 71.55 | 68.01 | 66.78 |
| 4711.37 | [Ar IV] | 5.90 | 6.62 | 6.59 | 6.85 |
| 4740.17 | [Ar IV] | 5.18 | 5.07 | 5.05 | 5.23 |
| 4861.33 | H I | 100 | 100 | 100 | 100 |
| 4958.91 | [O III] | 346.61 | 317.16 | 323.08 | 309.97 |
| 5006.84 | [O III] | 1012.97 | 940.22 | 941.44 | 917.49 |
| 5411.52 | He II | 3.34 | 6.28 | 5.93 | 5.88 |
| 5875.66 | He I | 10.08 | 9.73 | 10.00 | 10.22 |
| 6300.30 | [O I] | 3.92 | 0.16 | 0.16 | 0.12 |
| 6312.10 | [S III] | 0.76 | 1.11 | 1.09 | 1.08 |
| 6363.77 | [O I] | 1.02 | 0.05 | 0.05 | 0.04 |
| 6527.11 | He II | 3.56 | 0.50 | 0.47 | 0.01 |
| 6548.10 | [N II] | 6.13 | 7.27 | 7.05 | 7.39 |
| 6562.77 | H I | 297.84 | 297.23 | 293.41 | 298.06 |
| 6583.50 | [N II] | 20.43 | 20.31 | 20.58 | 20.51 |
| 6678.16 | He I | 3.14 | 2.54 | 2.58 | 2.69 |
| 6716.44 | [S II] | 1.49 | 1.27 | 1.27 | 1.14 |
| 6730.82 | [S II] | 2.04 | 1.81 | 1.80 | 1.60 |
| 6890.90 | He II | 0.48 | 0.75 | 0.7 | 0.71 |
| 7005.67 | [Ar V] | 0.85 | 3.15 | 3.14 | 3.36 |
| 7065.25 | He I | 2.96 | 4.28 | 4.34 | 4.53 |
| 7135.80 | [Ar III] | 9.05 | 10.89 | 10.80 | 10.76 |
| 7319.45 | [O II]b | 1.82 | 1.82 | 1.78 | 1.45 |
| 7330.20 | [O II]b | 1.79 | 0.98 | 0.96 | 0.78 |
| 7751.06 | [Ar III] | 2.67 | 2.61 | 2.58 | 2.58 |
| 8045.63 | [Cl IV] | 1.39 | 1.84 | 1.79 | 1.92 |
| 8237.15 | He I | 1.85 | 1.79 | 1.67 | 1.69 |
| F(25 μm) | dust cont. | 4.14Jy | 3.83Jy | 3.87Jy | 4.10Jy |

Table 4
Elemental abundances of IC 2003 derived from empirical methods and different models.

| Abundances | NEAT | KMH_nh | MRN_h | KMH_h |
| --- | --- | --- | --- | --- |
| He | 0.12 | 0.2 | 0.15 | 0.15 |
| N | $4.50 \times 10^{-5}$ | $4.63 \times 10^{-5}$ | $5.46 \times 10^{-5}$ | $5.69 \times 10^{-5}$ |
| O | $1.95 \times 10^{-4}$ | $1.66 \times 10^{-4}$ | $1.63 \times 10^{-4}$ | $1.57 \times 10^{-4}$ |
| C | $1.79 \times 10^{-4}$ | $2.67 \times 10^{-4}$ | $2.85 \times 10^{-4}$ | $2.96 \times 10^{-4}$ |
| Ne | $7.60 \times 10^{-5}$ | $1.20 \times 10^{-4}$ | $1.20 \times 10^{-4}$ | $1.20 \times 10^{-4}$ |
| S | $1.80 \times 10^{-6}$ | $3.59 \times 10^{-5}$ | $3.59 \times 10^{-5}$ | $3.59 \times 10^{-5}$ |
| Ar | $1.08 \times 10^{-6}$ | $1.01 \times 10^{-6}$ | $1.01 \times 10^{-6}$ | $1.01 \times 10^{-6}$ |
| Cl | $5.73 \times 10^{-8}$ | $3.60 \times 10^{-9}$ | $3.60 \times 10^{-9}$ | $3.60 \times 10^{-9}$ |
| N/O | 0.231 | 0.28 | 0.33 | 0.36 |
| C/O | 0.92 | 1.61 | 1.75 | 1.88 |

The electron temperature variation with the radial distance from the inner edge of the nebula is plotted in Fig. 1., which shows a significant difference between the curves of model 1 and model 2. The largest enhancement of electron temperature near the central star, when compared with other models, indicates that the grain heating effect is enhanced in the inner region of the PN. Dust temperature and the dust-to-gas mass ratio for model 3 are the same, with their respective values in model 1 and model 2, as seen in Table 3. The ionization structures of helium, carbon, nitrogen, and oxygen at their different ionization stages are shown in Fig. 2, which depicts the radial location at which these ions are seen. This figure also shows that the Lyman photons required to ionize hydrogen are trapped inside the nebula, indicating that IC 2003 is an ionisation-bounded nebula.(see Fig. 3)

### 4.3.1. Nebular parameters

Table 4 lists the values of nebular abundances derived from model 3, compared with their respective values of model 1 and model 2 and with the empirically calculated values. The elemental abundances derived from model 3 differ significantly from their values in model 1 and model 2, depicting that accurate abundance determination in a photo-ionized nebula requires to incorporate the energy input from grain photo-electric heating effectively. Particularly, the N/O and C/O ratios are found to be highest in model 3 when compared to the other two models and the values derived from empirical methods as well. The higher N/O ratio of model 3 (0.36) indicates that IC 2003 is a Type II PN (Peimbert, 1978) with a low mass progenitor, and the C/O ratio of 1.88 implies that it has a very strong carbon-rich circum-stellar environment. Nebular parameters such as $R_{in}$ is lower for model 3 when compared to the non-heating models, and the filling factor is larger as compared to other models, as seen in Table 2. The dust temperature $T_d$ and $m_d/m_g$ for all three models are the same. As model 3 takes into account the processes required to determine the nebular thermal and ionization structures in detail and also considers the more realistic model atmosphere for the H-poor central stars, the derived parameters of IC 2003 are more accurate than the values estimated earlier for this PN (Kingsburgh and Barlow, 1994; Wesson et al., 2005).

The KMH heating model explains the large negative temperature gradient observed particularly in [WR]PNe due to grain heating, as seen in Fig. 2. In comparison with other models, it shows that the abundance derivation in PNe, which traces the AGB nucleosynthesis, is strongly related to the temperature gradient within the ionized nebula. Also, our study brings out that the grain size distribution in this PN can be better described by the KMH grain size distribution rather than the MRN, which is usually adopted for the ISM. This could be true for many other PNe, which requires a detailed study similar to the one presented here for IC 2003.

### 4.3.2. Central star parameters

The values of central star parameters derived from model 3 are presented in Table 4 and are compared with their respective values of model 1 and model 2. The central star parameters $T_{eff}$, $L_*$ for all the models presented here are significantly higher than their respective values in the black-body model of MAS22 (155kK and $1010L_\odot$). As the model atmosphere has many metallic lines absorption, the temperature has to be higher than the Plankian value to give the required far-UV flux. This is seen from the SED of the central star for the black body and the model atmosphere for the same $T_{eff}$ in Fig. 1. We have obtained the stellar model tracks for the H-poor CSPNe from the grids of evolutionary tracks given by LA PLATA Stellar Evolution and Pulsation Research Group for PG1159 pre-white dwarfs (Miller Bertolami and Althaus, 2006) (available at http://evolgroup.fcaglp.unlp.edu.ar/TRACKS/tracks.html). This was taken for four different progenitor masses, viz. 2.2, 2.5, 3.05 and 3.5 $M_\odot$ are plotted in the HR diagram in Fig. 4, and the position of the central star with its parameters ($T_{eff}$, $L_*$) of model 3 is marked with an asterisk in this figure. We interpolated the tracks between 3.05 and 3.5 $M_\odot$ models, which passes through the asterisk to calculate the progenitor mass of IC 2003. It is lying to be at ∼ 3.26 $M_\odot$ which corresponds to a CSPN mass of ∼ 0.636 $M_\odot$. We also verified the calculated mass using the relation between log(N/O) and CSPN mass given by Maciel et al. (2010), which shows a CSPN mass of ∼ 0.646 $M_\odot$ for the N/O ratio derived from model 3. Keeping other parameters fixed, $L_*$ and $T_*$ were varied around their derived values in model 3 until $CLOUDY$ could sense this variation and differentiate the corresponding models with their $\chi^2$ values. From this, the maximum possible deviation in the parameters $T_{eff}$ and $L_*$ arrived at ± 2000 K and ± 200 $L_\odot$ respectively, which is shown as a plus sign in Fig. 4.

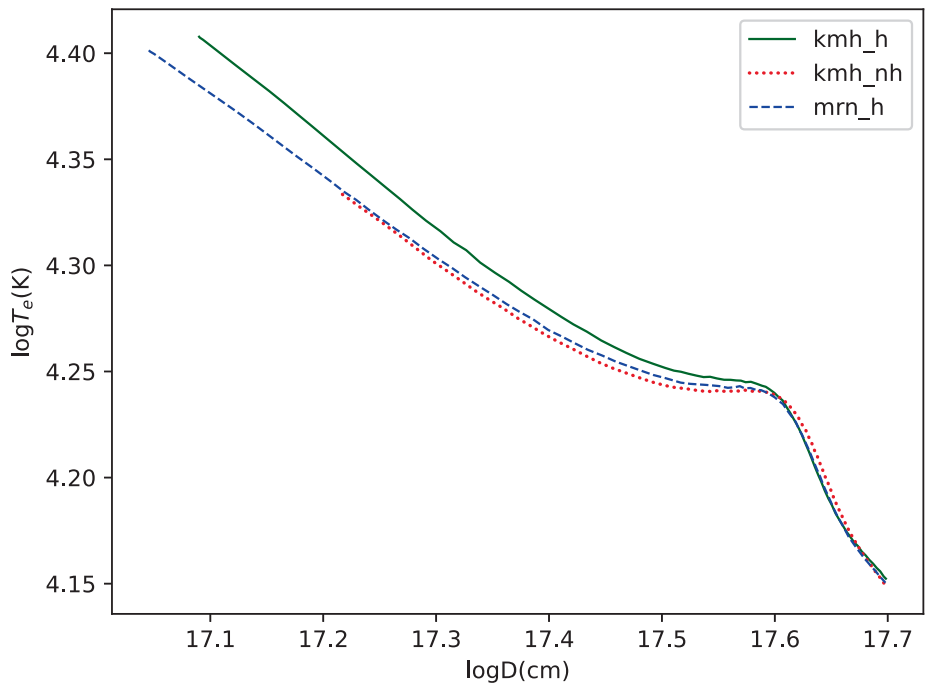

Fig. 2. Radially outward electron temperature profile of IC 2003 for different models.

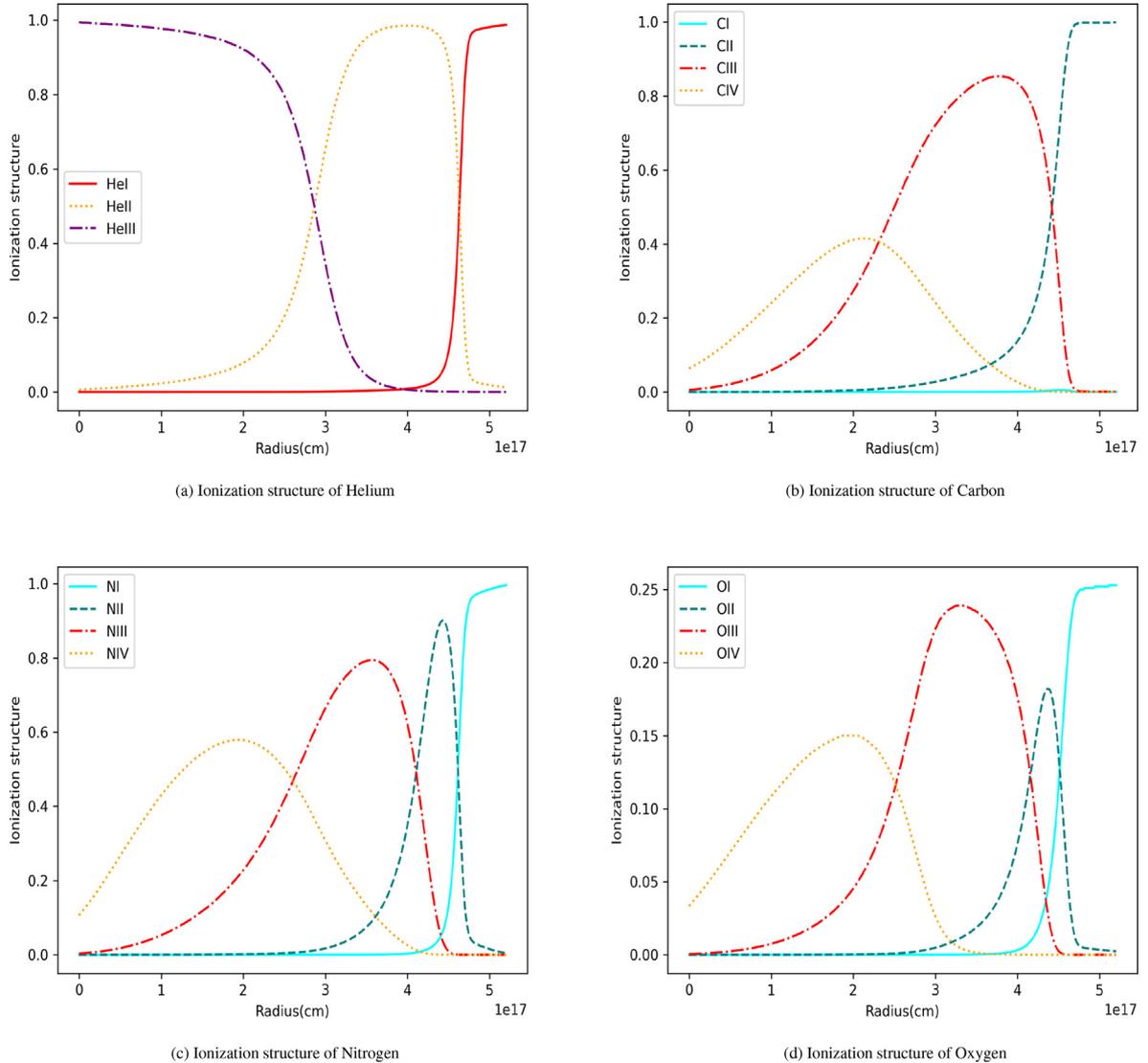

Fig. 3. Ionization structures of different elements at their different ionization stages (see the inset at each plot).

## 5. Conclusion

We study the thermal and ionization structure of PN IC 2003 using the 1-D Dusty photo-ionization model $CLOUDY$ and show that dusty photo-ionization modelling play key roles in unravelling the physical and chemical structures of [WR]PNe. We address the importance of grain heating of the nebula for MRN and KMH size distributions with and without quantum heating. We constrain these models using UV, optical spectroscopy and $IRAS$ photometry. Such a study provides a significantly better understanding of the PN than one arrives using the empirical abundances. The key findings from our study are:

1. The effect of photo-electric heating affects major cooling emission lines in the UV regime. The emission line flux

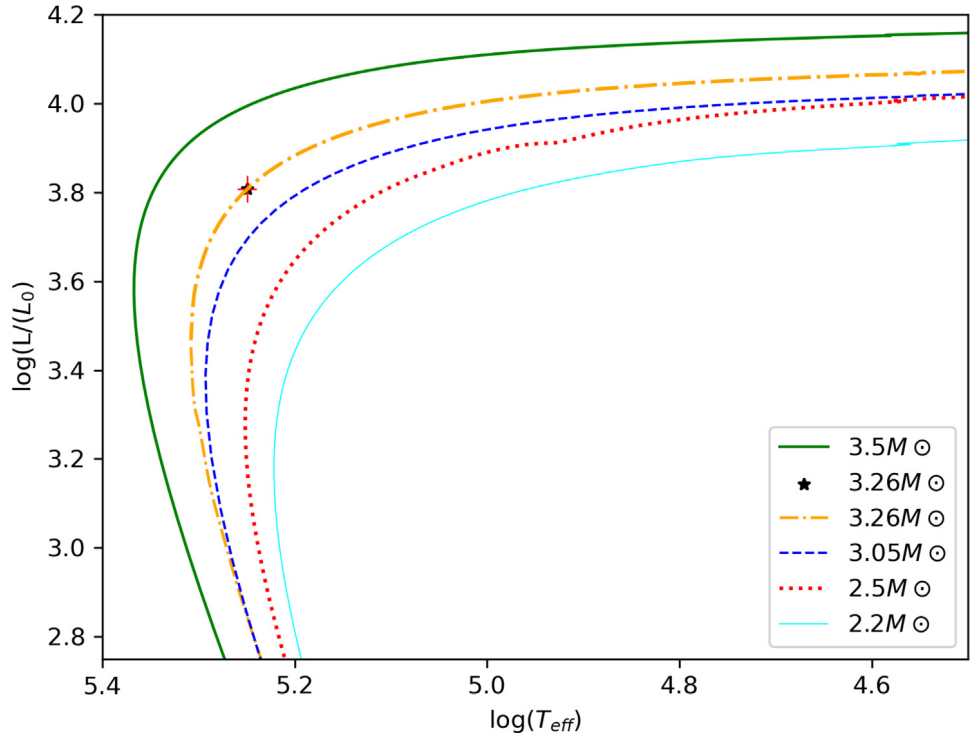

Fig. 4. Evolutionary tracks for H-poor central star planetary nebula for different progenitor masses 2.2, 2.5, 3.05 and 3.5 $M_\odot$.

of [CIV] 1549 Å and of [CIII] 1909 Å are significantly enhanced by including the quantum photo-electric heating by small dust grains.

2. The H-poor stellar model atmosphere using the Rauch spectral energy distribution grids has a significant impact on the nebular spectrum, and the black body approximation doesn't produce the realistic emission strengths of [WR] CSPNe in the UV to extreme UV region.

3. Lyman photons ionizing hydrogen are trapped within the nebula, suggesting that IC 2003 is an ionization-bounded nebula.

4. Dust heating with the KMH dust size distribution model fits well with the observations. The grain-gas heating ratio for this model is significantly higher than the value for the MRN heating model. This shows that grain size distributions follow KMH rather than MRN.

5. The derived nebular abundances and the N/O and C/O ratios for the KMH heating model differ significantly from their empirical values and the values from other models. The derived luminosity and temperature of the central star were used to place the central star in the HR diagram along with the evolutionary tracks of H-poor CSPNe, which suggests a CSPN mass of 0.636 $M_\odot$, which corresponds to a progenitor mass of 3.26 $M_\odot$ for PN IC 2003.

**Declaration of Competing Interest**

The authors declare that they have no known competing financial interests or personal relationships that could have appeared to influence the work reported in this paper.

**Acknowledgments**

This research has made use of the NASA/IPAC Infrared Science Archive, which is funded by the National Aeronautics and Space Administration and operated by the California Institute of Technology. The research work was supported by the DST-SERB grant of Govt. of India (Grant No: CRG/2020/000755) received by C. Muthumariappan. We thank Dr. M. M. Miller Bertolami for his suggestions on the stellar evolutionary tracks. The authors thank the VBT observing staff for their help in taking the optical spectroscopic observations. The authors thank Mr. G. Selvakumar for the valuable discussion on data analysis.